\documentclass[12pt]{article}
\usepackage{epsfig}
\title{ \bf Thermal QCD Sum Rules for $\sigma (600)$ Meson}

\author{El\c{s}en Veli Veliev *,  Takhmassib M. Aliev **
\\ ** Physics Department, Kocaeli University, Umuttepe Yerle\c{s}kesi \\
41380 Izmit, Turkey \\ e-mail: elsen@kocaeli.edu.tr
\\ *** Physics Department, Middle East Technical University, \\
06531 Ankara, Turkey \\
e-mail: taliev@metu.edu.tr}
\date{}
\begin{document}
\setlength{\baselineskip}{24pt}
\maketitle
\setlength{\baselineskip}{7mm}
\begin{abstract}
In the present work, the temperature dependence of the scalar mesons
parameters is investigated in the framework of thermal QCD sum
rules. We calculate  $\sigma$ -pole and the non-resonant two-pion
continuum contributions to the spectral density. Taking into account
additional operators appearing  at finite temperature, the thermal
QCD sum rules are derived. The temperature dependence of the shifts
in the mass and leptonic decay constant of scalar $\sigma (600)$
meson is calculated.
\end{abstract}

\setcounter{page}{1}
\section{Introduction}
The QCD sum rules method \cite{1}, proposed about three decades ago,
is one of the powerful methods for investigating the properties of
hadrons. This method has been extensively used as an efficient tool
to study the masses, decay form factors and so on \cite{2}. In
recent years there has been increasing interest in the modification
of hadronic properties at finite temperature in order to understand
the results of the heavy ion collision experiments.

The QCD sum rules method is extended to the finite temperature in
\cite{3} and finite temperature sum rules  have several new
features. One of them is the interaction of the current with the
particles of the medium. This effect requires modifying hadron
spectral function. The other novel feature is the breakdown of
Lorentz invariance by the choice of reference frame
\cite{4}-\cite{6}.  Due to the residual O(3) symmetry more operators
with the same dimension appear in the operator product expansion
(OPE) at finite temperature compared to those at zero temperature.
Taking into account both complications, the investigation of OPE for
thermal correlator of the two vector currents,  and the thermal QCD
sum rules for vector mesons have been realized in \cite{7} and
\cite{8}, respectively. Also, nuclear medium modifications of meson
parameters are widely discussed in the literature
\cite{9}-\cite{10}.

In the present work, we investigate the properties of the scalar
$\sigma$ meson in the framework of thermal QCD sum rules. The basic
idea of thermal QCD sum rules is to get information about
temperature dependence of hadron parameters, studying the same
correlator, both at high temperature where the quark-gluon plasma is
expected and at low temperature, where the hadronic phase is
dominated. Note that the nature of light scalar mesons is still an
open problem and is the subject of intensive and continuous
theoretical \cite{11} and experimental investigations \cite{12}. Can
we get any new information about the nature of the scalar mesons
from the thermal QCD analysis? Present work is addressed to the
investigation of this problem.

The paper is organized as follows. In section 2 we derive the
thermal QCD sum rules for scalar $\sigma (600)$ meson. In section 3
we present our numerical calculations. This section also contains
discussion and our conclusion.
\section{Thermal QCD sum rules for scalar sigma mesons}
In this section we construct the thermal sum rules for scalar
$\sigma (600)$ meson. For this purpose we consider the thermal
average of correlation function
\begin{equation}\label{eqn1}
T(q)=i \int d^{4}x e^{iq\cdot x} \langle T(J(x)J(0))\rangle, \\
\end{equation}
where $J(x)=\frac{1}{\sqrt{2}}(\overline{u}u+\overline{d}d)$ is the
interpolating current with the $\sigma$ meson quantum numbers.  The
thermal average of any operator   is determined by following
expression
\begin{equation}\label{eqn2}
\langle O\rangle=Tr e^{-\beta H}O/Tr e^{-\beta H}, \\
\end{equation}
where $H$ is the QCD Hamiltonian, and  $\beta=1/T$ stands for the
inverse of the temperature $T$  and traces are carried out over any
complete set of states.

The fundamental assumption of Wilson expansion is that the product
of operators at different points can be expanded as the sum of local
operators with momentum dependent coefficients in the form:
\begin{equation}\label{eqn3}
T(q)=\sum_n C_{n}(q^2)\langle O_{n}\rangle , \\
\end{equation}
where $C_{n}(q^2)$ are called Wilson coefficients, and $O_{n}$ are a
set of local operators. In this expansion, the operators are ordered
according to their dimension $d$ . The lowest dimension operator
with $d=0$ is the unit operator associated with the perturbative
contribution. In the vacuum sum rules operators with dimensions
$d=3$  and $d=4$ composed of quark and gluon fields are the quark
condensate $\langle \overline{\psi}\psi \rangle$ and the gluon
condensate $\langle G^{a}_{\mu\nu}G^{a\mu\nu}\rangle$, respectively.
At finite temperature Lorentz invariance is broken by the choice of
a preferred frame of reference, and therefore new operators appear
in the Wilson expansion. In order to restore Lorentz invariance in
thermal field theory, four-vector velocity of the medium $u^{\mu}$
is introduced. Using four-vector velocity and quark/gluon fields, we
can construct a new set of low dimension operators $\langle
u\Theta^{f}u\rangle$ and $\langle u\Theta^{g}u\rangle$ with
dimension $d=4$ , where $\Theta^{f}_{\mu\nu}$ and
$\Theta^{g}_{\mu\nu}$ are fermionic and gluonic parts of energy
momentum tensor $\Theta_{\mu\nu}$, respectively. So, we can write
thermal correlation function in terms of operators up to dimension
four:
\begin{equation}\label{eqn4}
T(q)=C_{1} I + C_{2}\langle \overline{\psi}\psi\rangle +
C_{3}\langle G^{a}_{\mu\nu}
G^{a\mu\nu}\rangle + C_{4}\langle u\Theta^{f}u\rangle + C_{5}\langle
u\Theta^{g}u\rangle   . \\
\end{equation}
The Wilson coefficients in Eq.(\ref{eqn4}) are calculated in
\cite{13}, and renormalization group improved expression of OPE
leads to the following result
\begin{eqnarray}\label{eqn5}
&&T(Q)=\frac{3}{
8\pi^2}Q^2\left(\gamma-\ln\frac{4\pi}{Q^2}\right)+\frac{3}{Q^2}m
\langle \overline{\psi}\psi\rangle+\frac{g^2}{32 \pi^2 Q^2}\langle
G^{a}_{\mu\nu}G^{a\mu\nu}\rangle
\nonumber \\
&&+\frac{4}{16+3n_{f}}\left(\frac{4(u\cdot
Q)^2}{Q^4}+\frac{1}{Q^2}\right) \left[\langle u \Theta u \rangle +
\lambda\left(Q^{2}\right)\left(\frac{16}{3}\langle
u\Theta^{f}u\rangle-\langle u\Theta^{g}u\rangle\right)\right],
\end{eqnarray}
where $Q$ is the Euclidean momentum,
$\lambda(Q^2)=\left[\alpha_{s}\left(\mu^{2}\right)
/\alpha_{s}\left(Q^{2}\right)\right]^{-\delta/b}$ and $\Theta_{\mu
\nu}=n_f \Theta^{f}_{\mu\nu}+\Theta^{g}_{\mu\nu}$. At one loop level
 the constants $\delta$ and $b$ are given by
\begin{equation}\label{eqn6}
\delta=\frac{2}{3}\left(\frac{16}{3}+ n_{f}\right)\,\,\,\,\,\, and
\,\,\,\,\,\, b=11-\frac{2}{3}n_{f} \,\,, \\
\end{equation}
where $n_{f}$ is quark flavours number. The spectral representation
for the correlation function in $q_{0}$ at fixed $|\textbf{q}|$ can
be written as \cite{8}
\begin{equation}\label{eqn7}
T\left(q_{0}^{2},|\textbf{q}|\right)= \int_{0}^{\infty}d{q'}_{0}^{2}
\frac{N\left(q_{0}^{'},|\textbf{q}|\right)}{{q'}_0^{2}+Q_0^2}+
subtraction \,\,\,\,\,\,terms \,\, , \\
\end{equation}
where
\begin{equation}\label{eqn8}
N\left(q_{0},|\textbf{q}|\right)= \frac{1}{\pi}
ImT\left(q_{0},|\textbf{q}|\right)\tanh\left(\beta
q_{0}/2\right)\,\,\,\,\,\, and \,\,\,\,\,\, Q_{0}^{2}=-q_{0}^{2} . \\
\end{equation}
Note that, the subtraction terms are removed by the Borel
transformation. For this reason in further discussion we omit these
terms. Equating spectral representation and OPE, and performing
Borel transformations with respect to  $Q_{0}^{2}$  from both sides
Eq.(\ref{eqn7}), we obtain the QCD sum rules
\begin{equation}\label{eqn9}
\int_{0}^{\infty}dq_{0}^{2}e^{-q_{0}^{2}/M^2}
N\left(q_{0},|\textbf{q}|\right)=e^{-|\textbf{q}|^{2}/M^2}
\left[\frac{3 M^4}{8 \pi^2}+\langle O_1  \rangle+ \left(1-\frac{4\textbf{q}^2}{3 M^2}\right)\langle O_2  \rangle\right]\,\, , \\
\end{equation}
 where $M$ is Borel parameter, $\langle O_1\rangle$ and $\langle O_2\rangle$
 are the non-perturbative contributions of higher dimensional operators,
\begin{equation}\label{eqn10}
\langle O_1\rangle=3m \langle \overline{\psi}\psi\rangle +
\frac{g^2}{32 \pi^2} \langle G^{a}_{\mu\nu}G^{a\mu\nu}\rangle \,\, ,\\
\end{equation}
\begin{equation}\label{eqn11}
\langle O_2\rangle=-\frac{12}{16+3n_{f}}\left[\langle u\Theta u
\rangle + \lambda (M^2)\left(\frac{16}{3}\langle u\Theta^{f} u
\rangle-\langle
u\Theta^{g} u \rangle\right)\right]\,\, .  \\
\end{equation}

Now we consider the phenomenological part of the correlation
function. We shall work below the critical temperature, where the
physical spectrum is saturated by hadrons  . In this case, similar
to the vacuum QCD sum rules, the dominant contribution to the
spectral density comes from $\sigma$  mesons. We also calculate the
contribution of the non-resonant two-pion continuum.

Let us calculate $\sigma$-pole contribution to the correlator. The
leptonic decay constant $\lambda_{\sigma}$ of the  $\sigma$-meson is
given by, $\langle 0|J(0)|\sigma\rangle =
m_{\sigma}\lambda_{\sigma}$, where $m_{\sigma}$ is $\sigma$-meson
mass. In thermal field theory, the parameters $m_{\sigma}$ and
$\lambda_{\sigma}$ must be replaced by their temperature dependent
values. The vacuum value of the leptonic decay constant is obtained
from two point QCD sum rules and $\lambda_{\sigma}=200MeV$
\cite{14}. The absorptive part of the correlation function is
calculated by using the following field-current identity
\begin{equation}\label{eqn12}
J(x)=m_{\sigma}\lambda_{\sigma}\sigma(x)\,\, , \\
\end{equation}
and $\sigma$-meson contribution to thermal correlator can be written
as
\begin{equation}\label{eqn13}
T(q)=i m_{\sigma}^{2}\lambda_{\sigma}^{2}D_{11}^{\sigma}(q)\,\, . \\
\end{equation}
Here $D_{11}^{\sigma}(q)= \int d^{4}x e^{iq\cdot x} \langle
T(\sigma(x)\sigma(0))\rangle$ is the time ordered product of two
$\sigma$-meson fields (11-component of the finite temperature scalar
field propagator with mass $m_{\sigma}$ in the real time formalism)
and has the following form \cite{15}-\cite{16}
\begin{equation}\label{eqn14}
D_{11}^{\sigma}(q)=\frac{i}{q^2- m_{\sigma}^{2}+i\varepsilon}+2\pi n
(\omega_{q})\delta(q^2-m_{\sigma}^{2})\,\, , \\
\end{equation}
where $n(\omega_{q})$ is the Bose distribution function,
$n(\omega_{q})=\left[exp(\beta \, \omega_{q})-1\right]^{-1}$ and
$\omega_{q}=\sqrt{\textbf{q}^2+m_{\sigma}^{2}}$. The imaginary part
of correlation function can be simply evaluated using the formula \,
$\frac{i}{x+i \varepsilon}=\pi \delta(x)+i
P\left(\frac{1}{x}\right)$, which leads to
\begin{equation}\label{eqn15}
Im T(q)=\pi m_{\sigma}^{2}\lambda_{\sigma}^{2}(2 n(\omega_q)+1)
\delta\left(q^2-m_{\sigma}^{2}\right)\,\, . \\
\end{equation}
With the help of  $\delta$-function we obtain the following result
for $\sigma$-pole contribution to the spectral function
\begin{equation}\label{eqn16}
N(q)=m_{\sigma}^{2}\lambda_{\sigma}^{2}
\delta\left(q^2-m_{\sigma}^{2}\right). \\
\end{equation}
In order to calculate  the dependence of $m_{\sigma}$ and
$\lambda_{\sigma}$ on temperature, we consider appropriate loop
diagrams. Let us calculate the $\pi\pi$-contribution to the
amplitudes, which describes the interaction of the current with the
particles in the medium. This contribution to the correlation
function can be written as
\begin{equation}\label{eqn17}
T(q)=i g_{\sigma}^{2} \int \frac{d^4 k}{(2 \pi)^4}
D_{11}^{\pi}(k)D_{11}^{\pi}(k-q) \,\, , \\
\end{equation}
where  $D_{11}^{\pi}(k)$  is the 11-component of the finite
temperature propagator for pions and $g_{\sigma}=2,0 \,GeV$
\cite{17}-\cite{18}. The integration over $k_0$ in Eq.(\ref{eqn17})
can be evaluated using the residue theorem. After integration and
some simplifications for the imaginary part of the correlation
function we obtain
\begin{eqnarray}\label{eqn18}
Im T(q)&=&\pi g_{\sigma}^{2} \int
\frac{d\textbf{k}}{(2\pi)^3}\frac{1}{4
\omega_{1}\omega_{2}}\left(\left(1+n_1\right)\left(1+n_2\right)+n_1
n_2\right)\left(\delta\left(q_0-\omega_{1}-\omega_{2}\right)+
\delta\left(q_0+\omega_{1}+\omega_{2}\right)\right)
\nonumber \\
&+&\left(\left(1+n_1\right)n_2+\left(1+n_2\right)n_1\right)
\left(\delta\left(q_0-\omega_{1}+\omega_{2}\right)+
\delta\left(q_0+\omega_{1}-\omega_{2}\right)\right) \,\, ,
\end{eqnarray}
where,
\begin{equation}\label{eqn19}
n_1=n(\omega_1)\,\, , \,\,\,\,\, n_2=n(\omega_2)\,\, , \,\,\,\,\,
\omega_1=\sqrt{\textbf{k}^{2}+m_{\pi}^{2}}\,\, , \,\,\,\,\,
\omega_2=\sqrt{(\textbf{k}-\textbf{q})^{2}+m_{\pi}^{2}} \,\,\, .  \\
\end{equation}
At  values $q_0=\omega_1+\omega_2$ and  $q_0=\omega_1-\omega_2$ the
terms involving the density distributions can be written as
\begin{equation}\label{eqn20}
\left[\left(1+n_1\right)\left(1+n_2\right)+n_1 n_2 \right] \tanh \left( \frac{\beta q_0}{2}\right )= \left( n_1+n_2+1\right) \,\, , \\
\end{equation}
\begin{equation}\label{eqn21}
\left[\left(1+n_1\right)n_2+\left(1+n_2\right)n_1 \right] \tanh
\left( \frac{\beta q_0}{2}\right )= \left(n_2-n_1 \right) \,\, ,
\end{equation}
respectively. As can be seen,  delta function
$\delta(q_0-\omega_1-\omega_2)$ in Eq.(\ref{eqn18}) gives the first
branch cut, $q^2\geq 4m_{\pi}^{2}$, which coincides with zero
temperature cut that describes the standard threshold for particle
decays. On the other hand, delta function
$\delta(q_0-\omega_1+\omega_2)$  in Eq.(\ref{eqn18}) shows that an
additional branch cut arises  at finite temperature, $q^2\leq 0$ ,
which corresponds  to particle absorption from the medium.
Therefore, delta functions $\delta(q_0-\omega_1-\omega_2)$  and
$\delta(q_0-\omega_1+\omega_2)$ in Eq.(\ref{eqn18}) contribute in
regions $q^2\geq 4m_{\pi}^{2}$ and $q^2\leq 0$, respectively. Taking
into account both contributions, the spectral function can be
written as
\begin{eqnarray}\label{eqn22}
N(q)&=&g_{\sigma}^{2} \int\frac{k^2 \sin\theta dk d\theta}{(2 \pi)^2
\,\,2\omega_1}\left[\left(n_1+n_2+1\right)\theta\left(q^2-4m_{\pi}^{2}\right)
+\left(n_2-n_1\right)\theta\left(-q^2\right)\right]
\nonumber \\
& &\times \delta \left( q^2-2 q_0
\omega_1+2|\textbf{k}||\textbf{q}|\cos \theta \right)\,\, .
\end{eqnarray}
The integration over angle $\theta$  in Eq.(\ref{eqn22}) can be
evaluated using the constraint $|\cos
\theta_{\textbf{q},\textbf{k}}|\leq1$ , which leads to following
inequality
\begin{equation}\label{eqn23}
\frac{|q^2-2q_0 \omega_1|}{2|\textbf{k}|
|\textbf{q}|}\leq1 \,\, .\\
\end{equation}
The solution of this inequality at values $q^2\geq 4m_{\pi}^{2}$
give us the integration range of $\omega_1$ as $\omega_{-} \leq
\omega\leq \omega_{+}$ , where
\begin{equation}\label{eqn24}
\omega_{\pm}=\frac{1}{2}(q_0\pm |\textbf{q}|v) \,\, , \\
\end{equation}
\begin{equation}\label{eqn25}
v(q^2)=\sqrt{1-4m_{\pi}^{2}/q^2} \,\, . \\
\end{equation}
At $q^2\leq 0$ the region of variation of $\omega_{1}$ must be
$\omega_{+} \leq \omega_{1}<\infty$. Finally, the thermal spectral
function can be written as
\begin{equation}\label{eqn26}
N(q)=\frac{g_{\sigma}^{2}}{2
|\textbf{q}|}\int^{\omega_{+}}_{\omega_{-}} \frac{d
\omega_1}{(2\pi)^2}\left(n_1+n_2+1\right)\theta
 \left(q^2-4 m_{\pi}^{2}\right)+\frac{g_{\sigma}^{2}}{2 |\textbf{q}|}\int^{\infty }_{\omega_{+}}
\frac{d \omega_1}{(2\pi)^2}\left(n_2-n_1\right)\theta
 \left(-q^2\right)\,\, . \\
\end{equation}
Changing the variable $\omega_1$  to $x$  given by
$\omega_1=\frac{1}{2}\left(q_0+|\textbf{q}|x\right) $  , we finally
get the two pion contribution to the spectral function as
\begin{equation}\label{eqn27}
N(q)\equiv{g_{\sigma}^{2}}\frac{v(q^2)}{8\pi^2}+N_1(q)=\frac{g_{\sigma}^{2}\,v(q^2)}{8\pi^2}+
\frac{g_{\sigma}^{2}}{8\pi^2}\int^{v}_{-v}dx\,\,
n\left(\frac{1}{2}\left(q_0+|\textbf{q}|x\right)\right)\,\, ,
\,\,\,\,\,\, q^2\geq 4m^2
\end{equation}
\begin{equation}\label{eqn28}
N(q)\equiv
N_2(q)=\frac{g_{\sigma}^{2}}{16\pi^2}\int^{\infty}_{v}dx\,\, \left [
n\left(\frac{1}{2}\left(|\textbf{q}|x-q_0\right)\right)-
n\left(\frac{1}{2}\left(|\textbf{q}|x+q_0\right)\right)\right]\,\, ,
\,\,\,\, q^2\leq 0
\end{equation}

The QCD sum rules are obtained by equating theoretical and
phenomenological parts of correlation function. Taking into account
of expressions Eq.(\ref{eqn27}) and Eq.(28) in Eq.(\ref{eqn9}) we
get
\begin{eqnarray}\label{eqn29}
&&m_{\sigma}^2(T)\lambda_{\sigma}^2(T)e^{-m_{\sigma}^2(T)/M^2}+
\frac{g_{\sigma}^{2}}{8 \pi^2}e^{|\textbf{q}|^2/M^2}
\int_{4m_{\pi}^2+|\textbf{q}|^2}^{\infty}dq_0^2 e^{-q_0^2/M^2} v
(q^2)
\nonumber \\
&+&e^{|\textbf{q}|^2/M^2}\left(\int_{4m_{\pi}^2+
\textbf{q}|^2}^{\infty}dq_0^2
e^{-q_0^2/M^2}N_1\left(q_0,|\textbf{q}|\right)+
\int_{0}^{|\textbf{q}|^2}dq_0^2
e^{-q_0^2/M^2}N_2\left(q_0,|\textbf{q}|\right)\right)
\nonumber \\
&=&\frac{3 M^4}{8 \pi^2}+\langle O_1
\rangle+\left(1-\frac{4\textbf{q}^2}{3 M^2}\right)\langle O_2
\rangle \, .
\end{eqnarray}
As  the temperature approaches to zero, the two terms in bracket go
to zero and the thermal average of the operators on the right become
the expectation values, recovering the vacuum sum rules \cite{14}.
In the limit $|\textbf{q}|\rightarrow 0$, the sum rule (29)
simplifies considerably. Finally we obtain that
\begin{equation}\label{eqn30}
m_{\sigma}^2(T)\lambda_{\sigma}^2(T)exp(-m_{\sigma}^2(T)/M^2)+I_0(M^2)
+I_1(M^2)=\frac{3 M^4}{8 \pi^2}+\langle O \rangle \,\, ,\\
\end{equation}
where
\begin{equation}\label{eqn30}
I_0(M^2)=\frac{g_{\sigma}^{2}}{8 \pi^2}\int^{\infty}_{4m_{\pi}^2}ds
\,\,
v(s)exp(-s/M^2) \,\, , \\
\end{equation}
\begin{equation}\label{eqn31}
I_1(M^2)=\frac{g_{\sigma}^{2}}{4 \pi^2}\int^{\infty}_{4m_{\pi}^2}ds
\,\,
v(s)\,\,n(\sqrt{s}/2)exp(-s/M^2) \,\, ,\\
\end{equation}
\begin{equation}\label{eqn311}
\langle O \rangle=\langle O_1 \rangle+\langle O_2 \rangle \,\, ,\\
\end{equation}
%
\section{Numerical analysis of the shifts in mass and  leptonic decay constant}

In this section we present our results for the temperature
dependence of the shifts in $\sigma$  meson mass and  leptonic decay
constant. By derivativing with respect to $1/M^2$ from both sides of
the sum rules (30), and making some transformations we obtain
\begin{equation}\label{eqn32}
m_{\sigma}^2(T)=\frac{m_{\sigma}^4 \lambda_{\sigma}^2
exp(-m_{\sigma}^2/M^2)-J_1(M^2)+\eta \overline{\langle O_3
\rangle}}{m_{\sigma}^2 \lambda_{\sigma}^2
exp(-m_{\sigma}^2/M^2)-I_1(M^2)+ \overline{\langle O\rangle} } \,\, , \\
\end{equation}
\begin{equation}\label{eqn33}
\lambda_{\sigma}^2(T)=\lambda_{\sigma}^2 \frac{m_{\sigma}^2
\lambda_{ \sigma}^2+\left(\overline{\langle O
\rangle}-I_1(M^2)\right) exp(m_{\sigma}^2/M^2)}{m_{\sigma}^2
\lambda_{
\sigma}^2+\left(\frac{1}{M^2}-\frac{1}{m_{\sigma}^2}\right)\left[
J_1(M^2)-\eta\overline{\langle O_3
\rangle}+m_{\sigma}^2\left(\overline{\langle O
\rangle}-I_1(M^2)\right)\right]exp(m_{\sigma}^2/M^2)} \,\, ,
\\
\end{equation}
where  the bar on the operators means subtractions of their vacuum
expectation values and
\begin{equation}\label{eqn34}
\eta(M^2)=\frac{\delta M^2}{ b \ln(M^2/\Lambda^2)}\,\, , \\
\end{equation}
\begin{equation}\label{eqn35}
J_1(M^2)=\frac{g_{\sigma}^{2}}{4 \pi^2} \int^{\infty}_{4 m_{\pi}^2}
ds\,\, s\,\,
\upsilon(s)\,\,n \left(\sqrt{s}/2\right)exp(-s/M^2) \,\, , \\
\end{equation}
\begin{equation}\label{eqn36}
\overline{\langle O_3
\rangle}=-\frac{12}{16+3n_{f}}\lambda(M^2)\left(\frac{16}{3}\langle
u\Theta^{f}u \rangle-\langle u\Theta^{g}u \rangle \right)\,\, . \\
\end{equation}
For the numerical analysis, let us list thermal average of operators
contributing to the QCD sum rules. The temperature dependence of
quark condensate is known from chiral perturbation theory
\cite{19}-\cite{20}
\begin{equation}\label{eqn37}
\langle \overline{\psi}\psi \rangle=\langle 0|\overline{\psi}\psi |0
\rangle \left[1-\frac{n_f^2-1}
{n_f}\frac{T^2}{12 F^2}+ O(T^4) \right]  , \\
\end{equation}
where $n_f$ is number of quark flavors and $F=0.088\,\, GeV$. The
low temperature expansion of the gluon condensate has been studied
in article \cite{21}
\begin{equation}\label{eqn38}
\frac{g^2}{4 \pi^2} \overline{\langle G^{a}_{\mu\nu} G^{a \mu \nu }
\rangle}=-\frac{8}{9} \left( \langle \Theta ^{\mu}_{\mu}\rangle +
\sum_f m_f \overline{\langle \overline{\psi}\psi
\rangle}\right) , \\
\end{equation}
where the trace of the total energy momentum tensor $\Theta
^{\mu}_{\mu}$ is given by $\langle\Theta
^{\mu}_{\mu}\rangle=\langle\Theta \rangle-3p$, and for two massless
quarks in the low temperature chiral perturbation limit the trace
has following form \cite{19}
\begin{equation}\label{eqn39}
\langle\Theta
^{\mu}_{\mu}\rangle=\frac{\pi^2}{270}\frac{T^8}{F_{\pi}^{4}}\ln
\frac{\Lambda_{p}}{T}+ O(T^{10})\,\, . \\
\end{equation}
Here $\langle\Theta \rangle$ is the total energy density and $p$ is
the pressure, whose expressions are known in the low  temperature
region \cite{20}. The pion decay constant has the value of
$F_{\pi}=0.093GeV$ and the logarithmic scale factor is
$\Lambda_{p}=0.275GeV$. We also use the fact that, the quark and
gluon energy densities at finite temperature can be expressed as
$n_f\langle\Theta^{f}\rangle=\langle\Theta^{g}\rangle=\frac{1}{2}\langle\Theta\rangle$,
which agrees both with the naive counting of the degrees of freedom
and empirical studies of the pion structure functions \cite{4},
\cite{8}.

\begin{figure}[h]
\centerline{\epsfig{figure=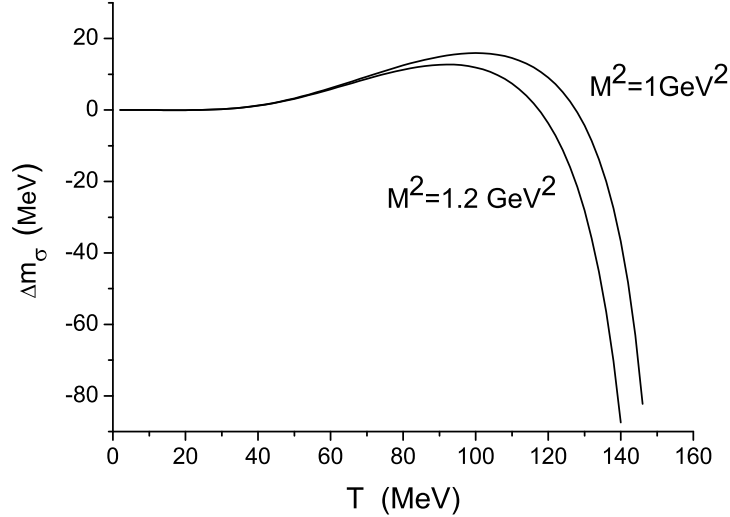,height=80mm}}\caption{ Shift
in the sigma meson mass as a function of temperature for $M^2=1\,\,
GeV^2$ and $M^2=1.2\,\, GeV^2$.} \label{BRmh0}
\end{figure}
For the numerical evolution of the above sum rule, we use the values
$m\langle\overline{\psi}\psi\rangle=-0.82\times10^{-4}\,\,GeV^{4} $,
$\Lambda=0.230\,\,GeV$ and $m_{\sigma}=0.6\,\,GeV$. We study the
dependence of $\sigma$  meson mass and leptonic decay constant on
$M^2$, when $M^2$ changes between $0.9\,\,GeV^{2}$ and
$1.4\,\,GeV^{2}$. This region of $M^2$ is obtained from the mass sum
rule analysis of the  $\sigma$ meson \cite{14}.

The shifts in $\sigma$ meson mass and leptonic decay constant as a
function of temperature for different values of $M^2$ is shown in
Fig.1 and Fig.2, respectively.
\begin{figure}[h]
\centerline{\epsfig{figure=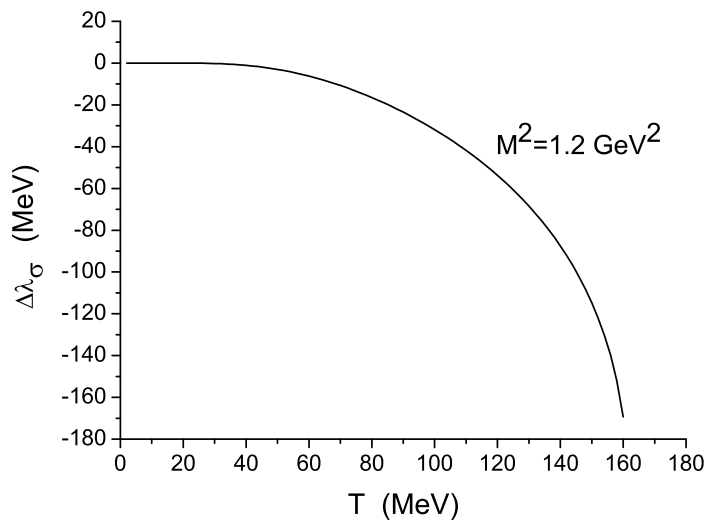,height=80mm}}\caption{ Shift
in the leptonic decay constant as a function of temperature for
$M^2=1.2\,\, GeV^2$.} \label{BRmh0}
\end{figure}
As seen, the results for $\Delta m_{\sigma}$ are stable for
temperatures up to $120\,\,MeV$. At high temperatures the results
for $\Delta m_{\sigma}$ becomes unstable and the contributions of
higher dimensional operators become important here, whose inclusions
might restore the stability in $M^2$ to higher temperatures. The
results for $\Delta \lambda_{\sigma}$ are stable and leptonic decay
constant decreases with increasing temperature and vanishes
approximately at temperature $T=160\,\,MeV$ . This situation may be
interpreted as a signal for deconfinement and agrees with
heavy-light mesons investigations \cite{23}. Numerical analysis
shows that the temperature dependence of $\Delta \lambda_{\sigma}$
is the same, when $M^2$ changes between $0.9\,\,GeV^{2}$ and
$1.4\,\,GeV^{2}$.

 Obtained results can be used for interpretation
heavy ion collision experiments. It is also essential to compare
these results with other model calculations. We believe these
studies to be of great importance for understanding phenomenological
and theoretical aspects of  thermal QCD.
\section{Acknowledgement}
This work is supported by the Scientific and Technological Research
Council of Turkey (TUBITAK), research project no.105T131, and the
Research Fund of Kocaeli University under grant no. 2004/4.
\end{document}